\documentclass[prl,floatfix,twocolumn,showpacs,amsmath,amssymb]{revtex4}
\usepackage{graphicx}
\usepackage{dcolumn}
\usepackage{bm}
\begin{document}

\title{From magnetism to one-dimensional spin liquid in the anisotropic 
triangular lattice}
\author{Dariush Heidarian,$^{1,2}$ Sandro Sorella,$^{1,2}$ and 
Federico Becca,$^{1,2}$}
\affiliation{
$^{1}$ International School for Advanced Studies (SISSA), Via Beirut 2, I-34014 Trieste, Italy \\
$^{2}$ CNR-INFM-Democritos National Simulation Centre, Trieste, Italy.
}

\date{\today} 

\begin{abstract}
We investigate the anisotropic triangular lattice that interpolates from 
decoupled one-dimensional chains to the isotropic triangular lattice and has 
been suggested to be relevant for various quasi-two-dimensional materials, 
such as Cs$_2$CuCl$_4$ or $\kappa$-(ET)$_2$Cu$_2$(CN)$_3$, an organic material
that shows intriguing magnetic properties. We obtain an excellent accuracy by 
means of a novel representation for the resonating valence bond wave function 
with both singlet and triplet pairing. This approach allows us to establish 
that the magnetic order is rapidly destroyed away from the pure triangular 
lattice and incommensurate spin correlations are short range. A non-magnetic 
spin liquid naturally emerges in a wide range of the phase diagram, with 
strong one-dimensional character. The relevance of the triplet pairing for 
$\kappa$-(ET)$_2$Cu$_2$(CN)$_3$ is also discussed.
\end{abstract}

\pacs{75.10.-b, 71.10.Pm,75.40.Mg}

\maketitle

When cooling down the temperature, the majority of materials undergo phase
transitions to ordered phases, that break some symmetry. Examples are 
ubiquitous in nature, e.g., magnets or superconductors, and define the 
paradigm in solid state physics. In the last years, a great effort has been 
done to determine and characterize new states of matter, which escape this 
conventional description. In this regard, one of the most intriguing case
is given by the so-called spin liquids, namely insulating phases that cannot 
be adiabatically connected to any band insulators.~\cite{misguich} The concept 
of spin liquid was introduced by Fazekas and Anderson~\cite{fazekas} and its
possible connection with the low-doping regime of high-temperature 
superconductors was highlighted by Anderson.~\cite{anderson} The standard 
picture of a spin liquid is given by the resonating valence bond (RVB) ansatz, 
a superposition of configurations in which couples of spins form singlets but 
change partner from one configuration to the other. After a long period 
dominated by the (wrong) prejudice that spin liquids cannot be actually 
stabilized, today there is an increasing evidence that they can be obtained 
in both microscopic models and real materials. 
Spin-liquid behavior has been suggested in various compounds: 
in two-dimensional (2D) triangular lattices,~\cite{coldea,kanoda} in Kagome 
materials,~\cite{mendels} and more recently in three-dimensional hyper-Kagome 
antiferromagnets.~\cite{takagi}

The 2D triangular lattice is the simplest structure in which the 
nearest-neighbor super-exchange leads to frustration. However, it is well 
proved that ideal Heisenberg spins with antiferromagnetic interactions on 
such a lattice display an ordered spin configuration, even for 
the spin-half case.~\cite{bernu,capriotti} Nevertheless, due to strong quantum 
fluctuations, the magnetic order parameter is highly reduced from its 
classical value~\cite{capriotti,chernyshev} and small perturbations may destroy
long-range order and drive the system towards a pure spin-liquid ground state. 
In this sense, a finite on-site repulsion $U$ (or equivalently multi-spin
interactions) may stabilize a magnetically disordered phase close to the 
metal-insulator transition.~\cite{motrunich,senechal} Another very interesting
possibility to further increase quantum fluctuations is to have different
super-exchange couplings along different spatial directions. This latter case
is particularly appealing because of its connection with various materials, 
such as Cs$_2$CuCl$_4$ and Cs$_2$CuBr$_4$~\cite{valenti} or a family of 
quasi-2D organic compounds.~\cite{mckenzie}

In this Letter, we consider a spin-half Heisenberg model defined on the 
anisotropic triangular lattice:
\begin{equation}\label{hamilt}
{\cal H} = J \sum_{(i,j)} {\bf S}_i \cdot {\bf S}_j +
J^\prime \sum_{\{i,j\}} {\bf S}_i \cdot {\bf S}_j,
\end{equation}
where ${\bf S}_i =(S^x_i,S^y_i,S^z_i)$ is the spin operator at the site $i$ 
and $(i,j)$ indicates nearest-neighbor sites along the ${\bf a}_1=(1,0)$ 
direction, while $\{i,j\}$ indicates nearest-neighbor sites along either 
${\bf a}_2=(1/2,\sqrt{3}/2)$ or ${\bf a}_3=(-1/2,\sqrt{3}/2)$.
Therefore, the model consists in one-dimensional (1D) chains coupled with 
zig-zag bonds $J^\prime$. Here, we consider clusters with $N=L^2$ sites 
and periodic boundary conditions along $L {\bf a}_1$ and $L {\bf a}_2$. 
Recent works showed a strong {\it one-dimensionalization}~\cite{ogata} 
and gapless $S=1/2$ excitations~\cite{yunoki} in a wide regime of frustration 
$J^\prime/J \lesssim 0.5$. The main limitation of these results is the 
inaccurate description of the magnetic correlations. In fact, works based upon 
series expansions~\cite{singh,pardini} showed that a magnetic spiral order may 
be present down to the 1D limit, with almost antiparallel spins along chains.
The fact that the 1D disordered phase is unstable towards the formation of 
incommensurate magnetic order has been also suggested by perturbative
approaches.~\cite{tsvelik,starykh} The situation is far from being clarified 
and more work is needed to understand the nature of the ground state.

\begin{figure}
\includegraphics[width=0.35\textwidth]{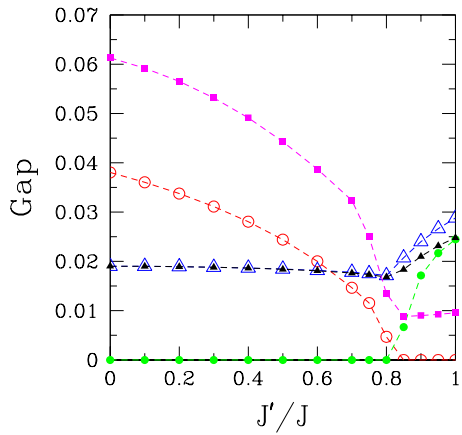}
\caption{\label{fig:gap6x6}
(Color on-line) Exact energy gap for the $6 \times 6$ cluster as a function of 
the frustrating ratio $J^\prime/J$. Full and empty circles indicate singlet 
states with $q=(0,0)$ and different reflection symmetry (the data show a level 
crossing for $J^\prime/J \sim 0.825$). Empty and full triangles indicate 
triplet excitations at $q=(\pi,\pi/3\sqrt{3})$ and $q=(\pi,\pi/\sqrt{3})$, 
respectively. Full squares indicate triplet excitations at $q=(3\pi/4,0)$.}
\end{figure}

Before considering our variational Monte Carlo calculations, it is useful to 
present exact results by the Lanczos method on the $6 \times 6$ cluster, 
see Fig.~\ref{fig:gap6x6}. As already obtained in Ref.~\onlinecite{bursill}, 
we find a level crossing in the ground state; this is due to a change,
around $J^\prime/J \sim 0.825$, in the quantum number of the reflection 
symmetry. In addition to ground-state properties, here we can also afford 
calculations for the important low-energy excited states. We find that the 
lowest triplet excitation has different quantum numbers for 
$J^\prime/J \gtrsim 0.775$, where $q=(3\pi/4,0)$, and for 
$J^\prime/J \lesssim 0.775$, where $q=(\pi,\pi/\sqrt{3})$. 
Remarkably, in a wide regime, the low-energy spectrum shows a clear 1D 
character with two (almost) degenerate triplet excitations with $q_x=\pi$, 
namely $q=(\pi,\pi/\sqrt{3})$ and $q=(\pi,\pi/3\sqrt{3})$, 
see Fig.~\ref{fig:gap6x6}.
 
\begin{figure}
\includegraphics[width=0.35\textwidth]{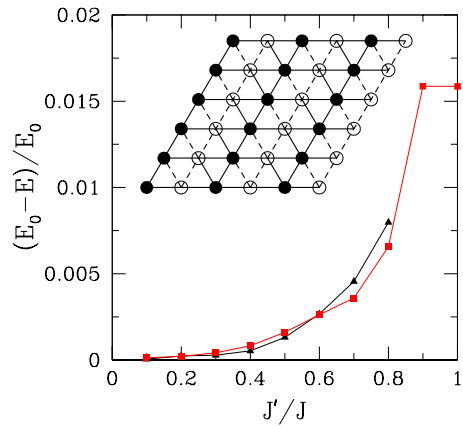}
\caption{\label{fig:accuracy}
(Color on-line) Accuracy for the energy on the $6 \times 6$ lattice, $E_0$ 
and $E$ denote the exact and the variational energies. The WF with 
decoupled chains (black triangles) and the one with $2 \times 1$ structure
and 120$^\circ$ order (red squares) are reported, see text for a detailed
description. The example of the $2 \times 1$ structure for the sign of the 
nearest-neighbor sites is also reported: solid and dashed lines denote 
positive and negative pairing amplitudes, respectively.}
\end{figure}

Let us now move to a detailed study of the Hamiltonian~(\ref{hamilt}) by using 
a variational wave function (WF) approach that includes both magnetic 
correlations and electronic pairing. In the original RVB approach the 
variational WF can be obtained by applying the Gutzwiller projector 
${\cal P}_G$ that completely suppress doubly occupied sites to the ground 
state of a mean-field BCS Hamiltonian.~\cite{anderson}
Within the same variational approach, a magnetic state is obtained by adding
an external field in the BCS Hamiltonian and considering a suitable 
long-range spin Jastrow factor that introduces the correct spin-wave 
fluctuations, i.e., ${\cal J}_s = 
\exp \left ( 1/2 \sum_{i,j} v_{ij} S^z_i S^z_j \right )$.~\cite{lugas}
The ground state of the mean-field Hamiltonian containing both electronic 
pairing and magnetism can be written in terms of a generalized complex pairing 
function $f^{\sigma_i,\sigma_j}_{i,j}$ that contains both singlet and
triplet components, so that the full variational WF is given by:~\cite{lugas}
\begin{equation} \label{psivmc}
|\Psi \rangle = {\cal J}_s {\cal P}_G \exp  \left \{ 
\frac{1}{2} \sum_{i,j,\sigma_i,\sigma_j}
f^{\sigma_i,\sigma_j}_{i,j} c_{i,\sigma_i}^{\dagger} c_{j,\sigma_j}^{\dagger}
\right \} |0 \rangle,
\end{equation}
where $c_{i,\sigma_i}^{\dagger}$ creates an electron with spin $\sigma_i$ on 
the site $i$. At present, all variational approaches on the lattice have 
optimized the WF by considering few {\it short-range} parameters of the BCS 
Hamiltonian (e.g., the BCS pairing and/or hopping amplitudes), implying a
{\it long-range} pairing function. Here, we generalize this variational 
approach without defining the mean-field BCS Hamiltonian. Instead, we directly 
optimize the pairing amplitude $f^{\sigma_i,\sigma_j}_{i,j}$. 
This approach allows us to have more variational freedom and, therefore,
provides a much less biased ansatz to the ground state. Let us now 
discuss the symmetries that we use for this quantity. First of all, we consider
independent $(\sigma_i,\sigma_j)$ values for ($\uparrow$,$\uparrow$), 
($\downarrow$,$\downarrow$), ($\uparrow$,$\downarrow$), and 
($\downarrow$,$\uparrow$) amplitudes. 
Then, in order to take into account magnetic correlations, we consider two 
different possibilities. The first one has a three-sublattice symmetry 
(suitable to the 120$^\circ$ order) and the second one has antiparallel spins 
along 1D chains and with two independent chains with different 
magnetic moment (suitable to describe the magnetic order for 
$J^\prime/J \ll 1$). Despite these limitations, the correlated WF of 
Eq.~(\ref{psivmc}), optimized in presence of the Jastrow factor ${\cal J}_s$, 
may show clear incommensurate spin-spin correlations, demonstrating that our 
approach is highly flexible and allows us to describe non-trivial spin 
correlations. In summary, for each bond and each spin case, we have three (two)
independent complex numbers for the first (second) case.
Finally, a $2 \times 1$ structure that implies an extra sign factor 
($+1$ or $-1$) for each $f^{\sigma_i,\sigma_j}_{i,j}$ is 
considered~\cite{yunoki2} for the case of three sublattices, 
see Fig.~\ref{fig:accuracy} for the sign convention of nearest-neighbor sites. 
Periodic or antiperiodic boundary conditions on $f^{\sigma_i,\sigma_j}_{i,j}$
are chosen, depending on $L$. Although the WF breaks the spin SU(2) symmetry, 
the actual value of the total spin square $\langle S^2 \rangle$ is as small 
as $0.07$ for the $18 \times 18$ cluster.
The fundamental ingredients are the presence of the $2 \times 1$ structure
for the signs of the pairing, relevant for $J^\prime \sim J$, and the direct 
optimization of $f^{\sigma_i,\sigma_j}_{i,j}$ (containing both singlet and
triplet components), which is afforded here for the first time. 
In the isotropic case, we obtain excellent results, which give an energy per 
site $E/J=-0.5470(1)$ in the thermodynamic limit, very close to our estimation 
of the exact value $E/J=-0.551(1)$ (which is extracted with the variance 
extrapolation of WFs with zero and one Lanczos step~\cite{sorella}) and much 
lower than previous estimates $E/J \sim -0.53$.~\cite{huse,weber}

\begin{figure}
\includegraphics[width=0.35\textwidth]{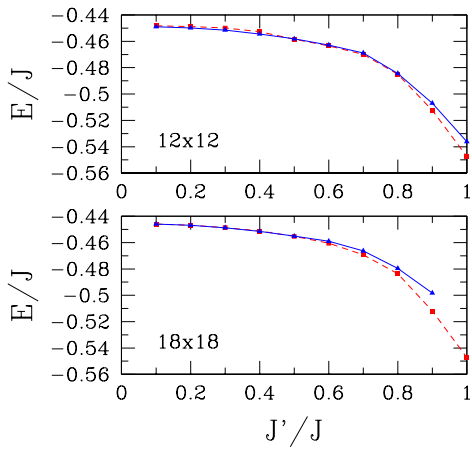}
\caption{\label{fig:energy}
(Color on-line) Energy per site for the $12 \times 12$ (upper panel) and the
$18 \times 18$ (bottom panel) for the WF with three-sublattice magnetization 
and $2 \times 1$ structure (red squares) and with two decoupled chains 
(blue triangles). See text for a detailed description of the WFs.}
\end{figure}

\begin{figure}
\includegraphics[width=0.45\textwidth]{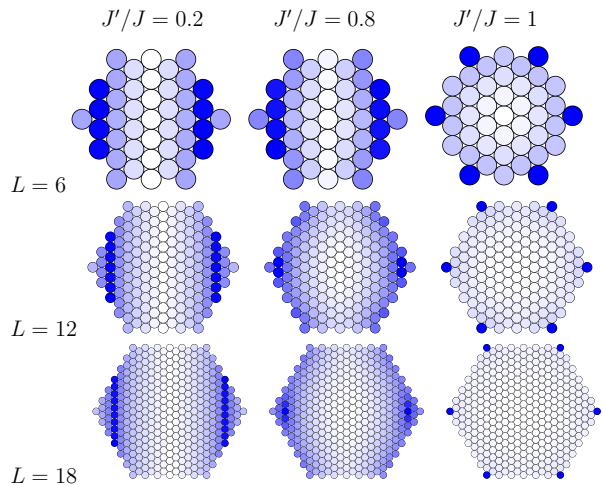}
\caption{\label{fig:Sq}
(Color non-line) Static spin-spin correlations $S(q)$ for different lattice 
sizes $N=L \times L$ and frustrating ratios $J^\prime/J$. A darker color 
indicates a bigger $S(q)$.}
\end{figure}

In Fig.~\ref{fig:accuracy}, we report the accuracy of the two WFs (with
three-sublattice or two-chain structure) for the $6 \times 6$ lattice. 
The full optimization of the pairing function allows us to reach a very good
accuracy in the whole range of our interest and, in particular, for 
$J^\prime/J \lesssim 0.5$. We notice that the level crossing present in 
exact calculations (see Fig.~\ref{fig:gap6x6}) is also present in the energy
of the two variational WFs, although it is shifted to $J^\prime/J \sim 0.6$.
In the case of a first-order transition, there is a macroscopic energy 
difference between the stable and the unstable states in both regions across
the transition point. However, by increasing the system size,
we observe that the two energies merge for small frustrating ratios, namely
$J^\prime/J \lesssim 0.6$, see Fig.~\ref{fig:energy}. This indicates that the 
transition becomes continuous in the thermodynamic limit. The tiny energy 
difference between the two WFs for $J^\prime/J \lesssim 0.6$ suggests an 
effective chain decoupling. Indeed, the two variational WFs are compatible
with a continuous transition: at the critical point, the two states coincide
and have vanishing inter-chain pairing amplitude.~\cite{note}

\begin{figure}
\includegraphics[width=0.35\textwidth]{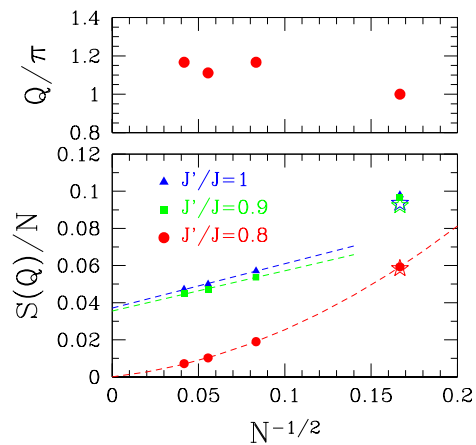}
\caption{\label{fig:Sqscaling}
(Color on-line) Lower panel: size scaling of the magnetic order parameter for 
different values of $J^\prime/J$. Stars indicate exact results for $N=36$ 
and lines are fits. Upper panel: position of the peak $q=(Q,0)$ for 
$J^\prime/J=0.8$.}
\end{figure}

Let us now move to the magnetic properties that can be assessed by the static 
spin-spin correlations:
\begin{equation}
S(q)=\frac{1}{N}\sum_{l,m} e^{i{\bf q} \cdot ({\bf R}_l-{\bf R}_m)} 
\langle {\bf S}_l \cdot {\bf S}_m \rangle.
\end{equation}
In Fig.~\ref{fig:Sq}, we show the results of $S(q)$ for three typical values 
of $J^\prime/J$ and three lattice sizes. In the isotropic case, we found that 
$S(q)$ has huge peaks at the corner of the Brillouin zone. 
The size scaling of $m^2=S(Q)/N$ with $Q=(4\pi/3,0)$ indicates a 
three-sublattice magnetic order, see Fig.~\ref{fig:Sqscaling}.
In the thermodynamic limit, we find $m^2 \sim 0.035$, which is larger
than $m^2 \sim 0.02$ found in previous works (within the present 
definition),~\cite{capriotti,chernyshev} showing that our approach favors
magnetic phases over spin liquids. Despite the fact that the magnetic
moment is considerably overestimated, the WF captures correct qualitative 
features. For $J^\prime/J=0.9$, we still obtain a finite value of $m^2$, which
is very close to the one found in the isotropic point. In this case, the 
peak of $S(q)$ stays at $Q=(4\pi/3,0)$, very close to the estimation given 
in Ref.~\onlinecite{pardini}. Moreover, another state can be stabilized, with 
incommensurate $Q$ but slightly higher energy. These facts indicate that the 
true incommensurability could be very small and it is not detectable with 
the available sizes. On the other hand, the size scaling at $J^\prime/J=0.8$ 
clearly indicates that $m^2 \to 0$ in the thermodynamic limit. 
Here, incommensurate spin correlations are found (see Fig.~\ref{fig:Sq}), 
demonstrating the flexibility of the variational WF. Furthermore, for 
$J^\prime/J \lesssim 0.6$, the spin-spin correlations display an almost 1D 
character: $S(q)$ does only depend upon $q_x$, whereas it has a flat behavior 
as a function of $q_y$, see Fig.~\ref{fig:Sq}. In this regime, the triplet 
components of the pairing amplitude are irrelevant, and we get a perfect RVB 
singlet state. Although we cannot exclude a tiny (incommensurate) magnetic 
order, as it was pointed out in Ref.~\onlinecite{tsvelik}, our calculations 
highlight the fact that the physical properties in the weakly coupled regime, 
i.e., $J^\prime/J \lesssim 0.6$, can be effectively represented as a 1D spin 
liquid down to very low energies (temperatures). On the other hand, for 
$J^\prime/J \gtrsim 0.6$ triplet components become fundamental to describe 
magnetic fluctuations. At the same time, for $J^\prime/J \gtrsim 0.6$ also the 
$2 \times 1$ structure of the pairing turns out to be important to gain 
energy, indicating a (second-order) transition between two spin 
liquids: one connected to the 1D case, having all equivalent sites,
and another one, having a $2 \times 1$ structure in the pairing 
function. No dimer order is found in the whole regime of frustration 
$0 \le J^\prime/J \le 1$.

\begin{figure}
\includegraphics[width=0.35\textwidth]{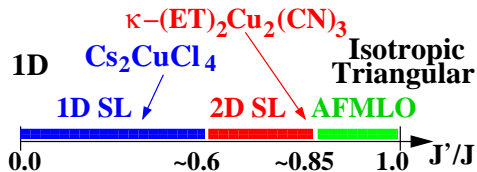}
\caption{\label{fig:phase}
(Color on-line) Phase diagram of the anisotropic triangular lattice, 
as obtained by our variational approach.}
\end{figure}

In summary, by using an improved variational approach, we have given strong 
evidence that a gapless spin liquid with negligible inter-chain coupling at 
low energy is stable over a wide region of the anisotropic triangular lattice. 
The complete phase diagram of this model (see Fig.~\ref{fig:phase}) can be 
worked out by considering both singlet and triplet pairing amplitudes that 
may give rise to magnetic order as well as incommensurate spin fluctuations. 
Our approach highlights the possibility to have two continuous transitions:
a first one between two spin liquids and another one from a 2D spin liquid and 
a magnetic phase. Close to the isotropic region, triplet correlations are 
particularly important, without necessarily implying magnetic order.
We finally remark that the existence of a non-magnetic state with explicit 
triplet pairing would naturally lead to a finite susceptibility at zero 
temperature and the onset of triplet superconductivity, in agreement with 
NMR experiments in $\kappa$-(ET)$_2$Cu$_2$(CN)$_3$.~\cite{kanoda}

F. Becca thanks R.R.P. Singh for useful discussions at KITP. 
We acknowledge support from CNR-INFM.


\begin{thebibliography}{99}
\bibitem{misguich} G. Misguich and C. Lhuillier,
   in ``Frustrated Spin Models'', Ed. H. T. Diep, World Scientific, New Jersey
   (2004).
\bibitem{fazekas}  P. Fazekas and P.W. Anderson, Philos. Mag. {\bf 30}, 
   423 (1974).
\bibitem{anderson} P.W. Anderson, Science {\bf 235}, 1196 (1987).
\bibitem{coldea} R. Coldea, D.A. Tennant, A.M. Tsvelik, and Z. Tylczynski,
   \prl {\bf 86}, 1335 (2001). 
\bibitem{kanoda} Y. Shimizu, K. Miyagawa, K. Kanoda, M. Maesato, 
   and G. Saito, \prl {\bf 91}, 107001 (2003); \prb {\bf 73}, 140407(R) (2006).
\bibitem{mendels} P. Mendels, F. Bert, M.A. de Vries, A. Olariu, A. Harrison, 
   F. Duc, J.C. Trombe, J.S. Lord, A. Amato, and C. Baines, \prl {\bf 98},
   077204 (2007).
\bibitem{takagi} Y. Okamoto, M. Nohara, H. Aruga-Katori, and H. Takagi, \prl
   {\bf 99}, 137207 (2007). 
\bibitem{bernu} B. Bernu, C. Lhuillier, and L. Pierre, \prl {\bf 69}, 2590 
   (1992).
\bibitem{capriotti} L. Capriotti, A.E. Trumper, and S. Sorella, \prl {\bf 82}, 
   3899 (1999).
\bibitem{chernyshev} S.R. White and A.L. Chernyshev, \prl {\bf 99}, 127004 
   (2007).
\bibitem{motrunich} O.I. Motrunich, \prb {\bf 72}, 045105 (2005).
\bibitem{senechal} P. Sahebsara and D. Senechal, \prl {\bf 100}, 136402 (2008).
\bibitem{valenti} K. Foyevtsova, Y. Zhang, H.O. Jeschke, and R. Valenti,
   arXiv:0812.2197.
\bibitem{mckenzie} For a review, see for instance, R.H. McKenzie, 
   Comments Cond. Matt. Phys. {\bf 18}, 309 (1998).
\bibitem{ogata} Y. Hayashi and M. Ogata, J. Phys. Soc. Jpn. {\bf 76} 053705 
   (2007).
\bibitem{yunoki} S. Yunoki and S. Sorella, \prl {\bf 92}, 157003 (2004).
\bibitem{yunoki2} S. Yunoki and S. Sorella, \prb {\bf 74}, 014408 (2006).
\bibitem{singh} Z. Weihong, R.H. McKenzie, and R.R.P. Singh, \prb {\bf 59},
   14367 (1999).
\bibitem{pardini} T. Pardini and R.R.P. Singh, \prb {\bf 77}, 214433 (2008).
\bibitem{tsvelik} M. Bocquet, F.H. Essler, A.M. Tsvelik, and A.O. Gogolin,
   \prb {\bf 64}, 094425 (2001).
\bibitem{starykh} O.A. Starykh and L. Balents, \prl {\bf 98}, 077205 (2007).
\bibitem{bursill} M.Q. Weng, D.N. Sheng, Z.Y. Weng, and R.J. Bursill, \prb 
   {\bf 74}, 012407 (2006).
\bibitem{lugas} M. Lugas, L. Spanu, F. Becca, and S. Sorella, \prb {\bf 74}, 
   165122 (2006).
\bibitem{sorella} S. Sorella, \prb {\bf 64}, 024512 (2001).
\bibitem{huse} D.A. Huse and V. Elser, \prl {\bf 60}, 2531 (1988).
\bibitem{weber} C. Weber, A. Lauchli, F. Mila, and T. Giamarchi, \prb {\bf 73},
   014519 (2006).
\bibitem{note} Residual 2D coupling between chains can be dynamically 
   generated by the Jastrow factor. 

\end{thebibliography}
\end{document}